\documentclass[10pt,twocolumn]{article} 
\usepackage{times}
\usepackage{cite}
\usepackage{graphicx}
\usepackage{amssymb}
\usepackage{amsmath}
\usepackage{url,hyperref}
\def\beq{\begin{equation}}
\def\eeq{\end{equation}}
\def\bea{\begin{eqnarray}}
\def\eea{\end{eqnarray}}
\usepackage{tikz}

\usepackage{graphicx}
\usepackage[font={small}]{caption}
\usepackage{xcolor}

\begin{document}

\title{Probing maximal zero textures with broken cyclic symmetry in inverse seesaw }

\author{Rome Samanta\footnote{rome.samanta@saha.ac.in}, Ambar Ghosal\footnote{ambar.ghosal@saha.ac.in}\\
\\
Saha Institute of Nuclear Physics, 1/AF Bidhannagar,
  Kolkata 700064, India
}

\maketitle
\thispagestyle{empty}

\begin{abstract}
   Within the framework of inverse seesaw mechanism we investigate neutrino mass matrices  invariant under cyclic  symmetry ($Z_3$) with maximal zero texture (6 zero textures). We explore two different approaches to obtain the cyclic symmetry invariant form of the constituent matrices. In the first one we consider explicit cyclic  symmetry  in the neutrino sector of the Lagrangian which dictates the emerged effective neutrino mass matrix ($m_\nu$) to be  symmetry invariant and hence leads to a degeneracy in masses. We then consider explicit breaking of the symmetry through a dimensionless parameter $\epsilon^{\prime}$ to remove the degeneracy. It is seen that the method doesn't support the current neutrino oscillation global fit data even after considering the correction from cyclic symmetry invariant charged lepton mass matrix ($m_l$) unless the breaking parameter is too large. In the second method, we assume  the same  forms of the neutrino mass matrices, however, symmetry is broken in the charged lepton sector. All the structures of the mass matrices are now  dictated by an effective residual symmetry of some larger symmetry group in the Lagrangian. For illustration, we exemplify a toy model based on softly broken $A_4$ symmetry group  which leads to one of the combination of $m_l$, $m_D$, $M_{RS}$ and  $\mu$  to generate effective $m_\nu$. All the emerged mass matrices predict a constraint range of the CP violating phases and atmospheric mixing angle along with an inverted hierarchical structure of the neutrino masses. Further,  significant predictions on  $\beta\beta0\nu$ decay parameter $|m_{11}|$ and the sum of the three light neutrino masses ($\Sigma_im_i$) are also obtained. 
\end{abstract}
\section{Introduction}
Although Type-I seesaw has been most popularly used to generate light neutrino Majorana masses, the high scale introduced through the incorporation of heavy right chiral $SU(2)_L \times U(1)_Y$ singlet ($\nu_{R}$) into this mechanism is beyond the reach of foreseeable collider experiments. Although the matter-antimatter asymmetry is naturally explained in this mechanism, lepton flavour violating process such as $\mu\rightarrow e\gamma$ are highly suppressed due to the new mass scale.  An alternative to this scenario is to consider inverse seesaw mechanism\cite{Mohapatra:1986bd,Bernabeu:1987gr,Mohapatra:1986aw,Schechter:1981cv,Schechter:1980gr,Fraser:2014yha,Hettmansperger:2011bt,Adhikary:2013mfa,Law:2013gma,Dev:2009aw,Hirsch:2009mx}  which contains additional left chiral $SU(2)_L \times U(1)_Y$ singlet field along with a low energy ($\sim$ keV) lepton number violating mass matrix $\mu$. The $9\times9$ neutrino mass matrix in this mechanism is written as 
 \bea
 M_\nu = \begin{pmatrix}
 0&m_D&0\\m_D^T&0&M_{RS}\\
 0&M_{RS}^T& \mu \\
 \end{pmatrix}
 \eea
 with the choice of the basis $(\nu_L,\nu_R^c,S_L)$. After diagonalization the low energy effective neutrino mass matrix comes out as
\bea
m_\nu &=& m_D M_{RS}^{-1}\mu ( m_D M_{RS}^{-1})^T \nonumber \\
 &=&F\mu F^T\label{e1} 
\eea
where $F=m_D M_{RS}^{-1}$. Here $m_D$ and $M_{RS}$ are Dirac type whereas $\mu$ is Majorana type mass matrix. The key feature in inverse seesaw is that the matrix $F$  plays an analogous role to that of $m_D$ in conventional Type-I seesaw. Now if one choose $m_D\sim $ 100 GeV and $M_{RS}\sim$ 10 TeV, the light neutrino masses will be $\simeq$ 0.1 eV which is protected by cosmology. Now due to the fewer number of experimental constraints a plausible approach is to minimize the number of parameters in the Lagrangian, in other words to consider maximal texture zeros\cite{Grimus:2004hf,Dev:2011jc,Frampton:2002yf,Whisnant:2014yca,Ludl:2014axa,Grimus:2013qea,Liao:2013saa,Fritzsch:2011qv,Merle:2006du,
Wang:2013woa,Wang:2014dka,Lavoura:2004tu,Kageyama:2002zw,Wang:2014vua,4zero1,Choubey:2008tb,
Chakraborty:2014hfa,4zero2,4zero3,4zero4,Ghosal:2015lwa,Samanta:2015hxa} in the fundamental mass matrices. Furthermore, since all the low energy neutrino parameters have yet not been fixed, a large number of studies are devoted on imposing flavour symmetry in the Lagrangian to minimize the number of parameters.  In the present work we adopt two different methods to incorporate the above ideas in the context of inverse seesaw mechanism. In the first method, to obtain viable neutrino mass matrix, we consider maximal zero textures along with a cyclic  symmetry ($Z_3$)\footnote{Motivation of considering cyclic  symmetry is that at the leading order one of the diagonalizing matrix is $U_{TBM}$.} between the generations of the neutrino fields motivated by the idea of Harrison, Perkins and Scott\cite{harri}. However, since cyclic symmetry gives rise to degenerate eigenvalues\cite{Adhikary:2013bma,Adhikary:2014qba,Pramanick:2013coa} it is necessary to lift the degeneracy and in the first method it is achieved through the minimal breaking of cyclic symmetry in $M_{RS}$ matrix. However, in this scheme the charged lepton correction is unable to satiate all the oscillation data simultaneously. Motivated by the first one (as the neutrino mass matrices are highly constrained with minimum number of parameters), in the second method, we concentrate on cyclic symmetry invariant form of the neutrino mass matrices rather inquire to the explicit  symmetry on the fields. Such type of form invariance can be realized through the residual symmetry\cite{Koide:2000zi,Damanik:2007cs,Koide:2006dn} of some bigger symmetry group in the Lagrangian. To realize one of the sub case, as a toy model we exemplify a softly broken $A_4$ group as the bigger symmetry group which is  completely broken in the charged lepton sector, though in the neutrino sector it breaks into $Z_3$ at the leading order. We particularly focus the cyclic symmetry ($Z_3$) invariant form of $m_D$ and $\mu$ assuming diagonal $M_{RS}$ in which residual $Z_3$ is broken. One of the combinations of the constituent matrices  leads to a bi-maximal type mixing\cite{Barger:1998ta,Elwood:1998kf,Ghosal:2000wg,Allanach:1998xi,1,2}, therefore, to generate nonzero $\theta_{13}$ we appeal to the  charged lepton mass matrix which is not $Z_3$ invariant as well as  in general  non diagonal\cite{Iizuka:2014mba,Kitabayashi:2013afa,Duarah:2012bd,Dev:2013esa}. It is  seen that the other emerged effective  $m_\nu$s also fail to explain the oscillation data unless charged lepton correction is considered.

\noindent
The plan of the present work is as follows:
Section \ref{s2} contains a qualitative description of the cyclic  symmetry and texture zeros for two different methods considered. A numerical estimation to obtain the viable parameter space satisfying the oscillation data is presented in Section \ref{s3}. Section \ref{s4} contains  summary of the present work.
\section{ Cyclic  symmetry and texture zeros}\label{s2}
\subsection{Explicit cyclic symmetry and texture zeros}
We assume the following cyclic symmetry in $\nu_{iL}$, $\nu_{iR}$ and $S_{iL}$ fields as 
\bea
\nu_{eL,R}\rightarrow \nu_{\mu L,R}\rightarrow\nu_{\tau L,R}\rightarrow\nu_{eL,R},\\
S_{eL}\rightarrow S_{\mu L}\rightarrow S_{\tau L}\rightarrow S_{eL}.
\eea
After imposition of the above cyclic  symmetry  general Dirac and Majorana type mass matrices look like 
\bea
m_D=\begin{pmatrix}
y_1&y_2&y_3\\
y_3&y_1&y_2\\
y_2&y_3&y_1
\end{pmatrix},
M=\begin{pmatrix}
x_1&x_2&x_2\\
x_2&x_1&x_2\\
x_2&x_2&x_1
\end{pmatrix}
.\eea
Now if we consider texture zeros along with the cyclic symmetry, clearly maximum number of zeros that can be accommodated  within the above matrices are 6. In Table \ref{t1} all the 6 zero textures of $m_D$ and $M_{RS}$ are presented. \\
\begin{table}[!h]
\caption{Texture zeros with cyclic  symmetry of $m_D$ and $M_{RS}$} \label{t1}
\begin{center}
\begin{tabular}{|p{7.6cm}|}
\hline
\multicolumn{1}{|c|}{{\bf $6$ zero textures of $m_D$ and $M_{RS}$}}\\
\hline
$m_{D}^1=\begin{pmatrix}
y_1&0&0\\
0&y_1&0\\
0&0&y_1\\
\end{pmatrix}$,
 $M_{RS}^1=\begin{pmatrix}
M_1&0&0\\
0&M_1&0\\
0&0&M_1\\
\end{pmatrix}$\\

$m_{D}^2=\begin{pmatrix}
0&y_2&0\\
0&0&y_2\\
y_2&0&0\\
\end{pmatrix}$,
$M_{RS}^2=\begin{pmatrix}
0&M_2&0\\
0&0&M_2\\
M_2&0&0\\
\end{pmatrix}$ \\
$m_{D}^3=\begin{pmatrix}
0&0&y_3\\
y_3&0&0\\
0&y_3&0\\
\end{pmatrix}$,
$M_{RS}^3=\begin{pmatrix}
0&0&M_3\\
M_3&0&0\\
0&M_3&0\\
\end{pmatrix}$\\
 \hline
\end{tabular}
\end{center}
\end{table} \\
Since the low energy lepton number violating mass matrix $\mu$ is Majorana type therefore, only one texture with 6 zeros is possible and is given as
\bea
\mu^1=diag(\mu_1,\mu_1,\mu_1)
.\eea  
Now, utilizing Eqn.(\ref{e1}) we construct $m_{\nu}$ and interestingly it is seen that along with diagonal $\mu$ any matrix  presented in Table \ref{t1} can not generate phenomenologically viable $m_{\nu}$, to be precise, all the emerged mass matrices ($m_\nu$) are diagonal. We now consider the next maximal texture zero (3 zero) structure of $\mu$, and is given by 
\bea
\mu^2=\begin{pmatrix}
0&\mu_2&\mu_2\\
\mu_2&0&\mu_2\\
\mu_2&\mu_2&0\\
\end{pmatrix}.
\eea
The above choice of $\mu$ matrix, along with the matrices presented in Table (\ref{t1}) enforces the $m_\nu$ to be nondiagonal. However, since the emerged $m_\nu$ is also cyclic  symmetry invariant and hence leading to a degeneracy in the eigenvalues, therefore removal of the degeneracy requires a small breaking of the symmetry. Since our philosophy is to find out a viable texture with least number of parameters, we consider minimal symmetry breaking in the different elements of $M_{RS}$ matrix only. For a compact view we present  Table \ref{t2} which contains all the combinations and the corresponding neutrino mass matrices ($m_\nu$)
\begin{table}[!h]
\caption{Different Composition of $m_D$ and $\mu$ matrices to generate $m_\nu$.}\label{t2}
\begin{center}
\begin{tabular}{|p{.8cm}|p{.8cm}|p{.8cm}|p{.8cm}|p{.8cm}|}
\cline{3-5}
\multicolumn{2}{c|}{$m_D$ and $\mu$}  & \multicolumn{1}{c|}{ $M_{RS}^{1\epsilon}$ }  & \multicolumn{1}{c|}{ $M_{RS}^{2\epsilon}$ }  & \multicolumn{1}{c|}{$M_{RS}^{3\epsilon}$}\\
\cline{3-5}
\multicolumn{2}{c|}{$\Updownarrow$}  & \multicolumn{3}{c|}{$m_\nu$ }\\ \hline
$m_D^1$&$\mu^2$ &$N^1$&$N^3$&$N^2$\\
\hline
$ m_D^2 $ &$\mu^2$ &$N^2$ & $N^1$ & $N^3$ \\
\hline
$m_D^3$ & $\mu^2$&$N^3$&$N^2$&$N^1$\\
\hline
$m_D^{1,2,3}$ & $\mu^1$ &$d^{1,2,3}$&$d^{3,1,2}$&$d^{2,3,1}$\\
\hline
\end{tabular}
\end{center}
\end{table}
with the definitions $M_{RS}^{1\epsilon}=diag(M_1+\epsilon,M_1,M_1) $, $M_{RS}^{2\epsilon}=diag(M_1,M_1+\epsilon,M_1) $, $M_{RS}^{3\epsilon}=diag(M_1,M_1,M_1+\epsilon) $. The $d^{i(i=1,2,3)}$s are some diagonal matrices of not our concern as those are obtained due to 6 zero texture of $\mu$ as discussed earlier. The matrices $N^1$, $N^2$ and $N^3$ arise due to 3 zero texture of $\mu$ matrix and explicitly their forms are given by
\bea
N^1=\begin{pmatrix}
0&A_1&A_1\\
A_1&0&B_1\\
A_1&B_1&0
\end{pmatrix},N^2=\begin{pmatrix}
0&B_2&A_2\\
B_2&0&A_2\\
A_2&A_2&0
\end{pmatrix}, \nonumber \\
N^3=\begin{pmatrix}
0&A_3&B_3\\
A_3&0&A_3\\
B_3&A_3&0
\end{pmatrix}\label{efmnu}
\eea
 with the definition of the parameters as 
\bea
A_i=\frac{\mu_2 y_i^2}{M_1(M_1+\epsilon)},
B_i=\frac{\mu_2 y_i^2}{M_1^2}.
\eea
$Phenomenological$ $consequences:$ As the left chiral neutrino fields obey cyclic symmetry, their charged lepton partners also follow the same. Hence, the charged lepton mass matrix ($m_l$) is diagonalized by trimaximal mixing matrix \cite{Koide:2000zi}. In the basis where the $m_l$ is diagonal the effective neutrino mass matrix will be modified by the trimaximal mixing matrix. However, it is found that due to the lack of sufficient number of parameters, all the mixing angles cannot be obtained simultaneously in their $3\sigma$ range. We also consider the  nondiagonal forms of $M_{RS}$ matrices (i.e., all the possible cases given in Table {\ref{t1}) and  find that the above conclusion is  valid for all the cases. Now at this stage one could  move one step ahead, i.e. one may consider three zero texture of $m_D$ and $M_{RS}$. In that case all the constraints from the oscillation data can be accommodated undoubtedly. However, in such a scenario as the effective number of parameters in the $m_\nu$ itself (without considering the charged lepton correction) increase, thus, the predictions on the light neutrino masses ($m_i$), their sum ($\Sigma_i m_i$) and neutrinoless double beta decay parameter ($|m_{11}|$) are less significant (vary in a wide range). Thus, since the maximality of zeros is our concern, in the next section we present an alternative approach to preserve the maximal zero textures of the constituent neutrino mass matrices. In this approach the required texture zero mass matrices with cyclic symmetry in the neutrino sector and  simple four zero textures with naturally broken $Z_3$ in the charged lepton sector are realized from an effective residual symmetry to reproduce the forms of $m_\nu$ matrices presented in Table \ref{t2} .\\
\subsection{Cyclic symmetry and texture zeros as an effective residual symmetry}\label{mwbr}
In this section we present a toy model based on $A_4$ symmetry as a bigger symmetry group. Due to spontaneous breaking of $A_4$, cyclic symmetry ($Z_3$) is preserved only in the neutrino sector while the charged lepton mass matrix is obtained with four zero Yukawa texture with decoupled third generation. Thus charged lepton correction also plays a crucial role to fit the extant data. However, before going into the detailed discussion, we would like to mention that although there are several cases in the analysis, we present a toy model only for one case. Furthermore, the symmetry group $A_4$ is not the only group to realize the cyclic symmetry with the texture zeros. Other symmetry groups such as $S_{4}$, $U(1)_{B-L}$ etc.\cite{Koide:2000zi,Ma:2014qra,Ma:2015mjd,Muramatsu:2016bda} can also lead to $Z_3$ invariance in the neutrino sector due to their spontaneous breaking. Now let us recall the problem we faced in the previous section. First, the maximal zero textures with cyclic symmetry in the neutrino sector do not entertain cyclic symmetry invariant form of the charged lepton mass matrix as far as the present experimental data is concerned. Apart from that one also needs to break cyclic symmetry in the neutrino sector since at the leading order it leads to a degeneracy in masses. Here, in the charged lepton sector, breaking of $Z_3$ is obtained due to spontaneous breaking of $A_4$ whereas in the neutrino sector the breaking scheme is similar to the previous section, i.e. the degeneracy is removed by due to a soft breaking term ($\epsilon$) in the elements of $M_{RS}$. Thus we need the structure of $M_{RS}$ due to minimal breaking as
\bea
M_{RS}={\rm diag }\hspace{1mm}(M_1,M_2,M_2)\label{mdgn}
\eea

with $M_1$ = $M_2+\epsilon$, to generate $N^{1,2,3}$ type mas matrices shown in Table \ref{t2}. Obviously such choice of $M_{RS}$ matrices with all nondegenerate eigenvalues are also consistent with the oscillation data. Although there are several effective $m_\nu$ arises due to suitable combinations of $m_D$, $\mu$ and $M_{RS}$, of them $N^{1}$ type matrix is a two parameter $\mu\tau $ symmetric matrix with zero diagonal entries. Consequently, the matrix leads to vanishing $\theta_{13}$ which is discarded by the present oscillation data at $>$ $10\sigma$ level \cite{th13}. Thus to generate nonzero $\theta_{13}$ corrections from the charged lepton sector \cite{Iizuka:2014mba,Kitabayashi:2013afa,Duarah:2012bd,Dev:2013esa,Liao:2015hya} should be taken into account. As a simplistic scenario, in this section we consider corrections from all the three sectors of $m_l$. These simple structures of $m_l$ are well motivated by popular discrete flavor groups which are used to explain neutrino mass and mixing. Here we consider $A_4$ as the flavor symmetry group. However, there are other groups, e.g. $S_4$ \cite{Shimizu:2015tta}, $Z_6$\cite{Dev:2013esa} etc. which can also lead to these structures of $m_l$. Interestingly, all the emerged $m_\nu$ which arises from $M_{RS}={\rm diag}\hspace{1mm}(M_1,M_2,M_3)$ also require charged lepton correction which we discuss in the next section.  Although there are several papers on $A_4$ symmetry we are motivated by Ref \cite{Hirsch:2009mx}.  We discuss the required  $A_4$ model in brief.

\begin{table}[!h]
\caption{Field content of the model with lepton and scalar assignment}\label{t3}
\begin{center}
\begin{tabular}{|p{1cm}|p{.6cm}|p{.6cm}|p{.6cm}|p{.6cm}|p{.6cm}|p{.6cm}|p{.6cm}|p{.6cm}|}
\cline{2-9}
\multicolumn{1}{c|}{}  & \multicolumn{1}{c|}{ $L$ }  & \multicolumn{1}{c|}{ $l_{lR}$ }  & \multicolumn{1}{c|}{$N_{lR}$ }&\multicolumn{1}{c|}{$S_{lL}$ }&\multicolumn{1}{c|}{$\xi_{ch},\phi_{ch}$ }&\multicolumn{1}{c|}{$\xi_D$ }&\multicolumn{1}{c|}{$\xi_{RS}$ }&\multicolumn{1}{c|}{$\phi_\mu$ }\\
 \hline
$SU(2)_L$&$2$ &$1$ &$1$ &$1$ &$2$ &$2$  &$1$&$1$ \\
$ Z_3$ &$\omega$ &$1$ & $\omega$ & $\omega^2$&$\omega$ &$1$&$\omega$&$\omega^2$\\
$ Z_2$ &$+$ &$+$ & $-$ & $+$&$+$ &$-$&$-$&$+$\\
$ A_4$ &$3$ &$3$ & $3$ & $3$&$1,3$ &$1$&$1$&$3$ \\
\hline
\end{tabular}
\end{center}
\end{table}

\noindent
Fermionic part of the Lagrangian consists of four part as shown below
\bea
\mathcal{L}^{A_4}_{mass}=\mathcal{L}_{ch}+\mathcal{L}_{Dirac}+\mathcal{L}_{RS}+\mathcal{L}_{ss}.
\eea
Explicitly  each term  is written as
\bea
\mathcal{L}^{A_4}_{mass}=Y_{ch}\bar{L}l_{R}(\phi_{ch}+\xi_{ch})+Y_{D}\bar{L}N_{R}\xi_D\nonumber\\ + Y_M \bar{S_{L}}N_R\xi_{RS} + Y_u\bar{S_{L}^C} S_{L}\phi_\mu + {\rm h.c}
\eea
 with the following choice of the alignment $\xi_{ch}\sim<v^{\xi}_{ch}>$, $\phi_{ch}\sim <0,0,v^{\phi}_{ch}>$, $\xi_{D}\sim<v^{\xi}_{D}>$, $\xi_{RS}\sim<v^{\xi}_{RS}>$ and $\phi_\mu \sim <v^{\phi}_{\mu},v^{\phi}_{\mu},v^{\phi}_{\mu}>$. With such choice of VEV  one can realize the charged lepton correction from $1-2$ sector and  the structures of $m_D^1$ and $\mu^2$ along with the structure of $M_{RS}$ as 
 \bea
 M_{RS}={\rm diag} (M,M,M).
 \eea 
Here we assume $A_4$ group is generated by two generators
\bea
S=\begin{pmatrix}
1&0&0\\0&-1&0\\0&0&-1
\end{pmatrix},\hspace{.5cm} T=\begin{pmatrix}
0&1&0\\0&0&1\\1&0&0
\end{pmatrix}.
\eea 
The three dimensional representation satisfy the product rule
\bea
3\times 3=1+1^{\prime}+1^{\prime \prime}+3_S+3_A
\eea
where
 \bea
 1&=&a_1b_1+a_2b_2+a_3b_3\\
 1^{\prime}&=&a_1b_1+\omega^2 a_2b_2+ \omega a_3b_3\\
 1^{\prime \prime}&=&a_1b_1+ \omega a_2b_2+ \omega^2 a_3b_3
\eea
and 
\bea
3_S=(\frac{a_2b_3+a_3b_2}{2},\frac{a_3b_1+a_1b_3}{2},\frac{a_1b_2+a_2b_1}{2})\\
3_A=(\frac{a_2b_3-a_3b_2}{2},\frac{a_3b_1-a_1b_3}{2},\frac{a_1b_2-a_2b_1}{2}).
\eea
Thus, $A_4$ is spontaneously  broken in the charged lepton sector such that there is no effective $Z_3$ symmetry, however, the neutrino sector enjoys an effective residual $Z_3$ symmetry. As previously mentioned, $Z_3$ in $M_{RS}$ should be broken, we consider soft $A_4$ breaking term in the Lagrangian which is well studied earlier \cite{Adhikary:2008au,Ge:2010js,Adhikary:2006wi}. We consider $\mathcal{L}^{soft }$ as
\bea
\mathcal{L}^{soft}=\epsilon_{\alpha \beta} \bar{S}_{\alpha L}N_{\beta R}
\eea
where $\epsilon_{\alpha \beta \hspace{1mm}(\alpha,\beta=1,2,3)}$ is a coupling constant with mass dimension one and the double indices do not mean the summation over the indices. The term contributes to the ($\alpha,\beta$) element of $M_{RS}$ and breaks the residual $Z_3$ symmetry. Now if we choose ($\alpha,\beta=1$) then the soft term contributes to (1,1) element of $M_{RS}$ which in turn generates $N^1$ type $m_\nu$ with $m_D^1$ and $\mu^2$. In the following two sections we present detailed analysis of all the emerged $m_\nu$.

 \noindent
\subsubsection{ Two degenerate eigenvalues of $M_{RS}$ }\label{sec2.2.1}
 The matrix of type $N^{1,2,3}$ can be realized by changing the nondegenerate value at three different diagonal entries of $M_{RS}$ matrix given in Eqn.(\ref{mdgn} along with $m_D^1$ and $\mu^2$. First we consider the $N^1$ matrix which is given by  
 \bea
N^1=m_\nu=\begin{pmatrix}
0&yp&yp\\
yp&0&y\\
yp&y&0
\end{pmatrix}\label{m1}
\eea
with $y=\mu_2 y_1^2/M_2^2$, $p=M_2/M_1$. The matrix of Eqn. (\ref{m1}) is diagonalized by the unitary mixing matrix $U_\nu$ given by
\bea
U_\nu=\begin{pmatrix}
c_{12}&s_{12}&0\\
-\frac{1}{\sqrt{2}}s_{12}&\frac{1}{\sqrt{2}}c_{12}&-\frac{1}{\sqrt{2}}\\
-\frac{1}{\sqrt{2}}s_{12}&\frac{1}{\sqrt{2}}c_{12}&\frac{1}{\sqrt{2}}
\end{pmatrix}
\eea
where
\bea c_{12}=\frac{\sqrt{1+\frac{1}{\sqrt{1+8p^2}}}}{\sqrt{2}}\nonumber
\eea
and\bea
s_{12}=\sqrt{\frac{1}{2}-\frac{1}{2\sqrt{1+8p^2}}}.
\eea
Interestingly, if $p\rightarrow \infty$ ($M_2>>M_1$) we can have the well known bi-maximal mixing of neutrino masses. The eigenvalues of $m_\nu$ are given by
\bea
-m_1&=&\frac{1}{2}(y-\sqrt{1+8p^2}y)\nonumber\\
m_2&=&\frac{1}{2}(y+\sqrt{1+8p^2}y)\nonumber\\
-m_3&=&-y
\eea
where $m_2>m_1>m_3$. Now defining $\Delta m_{sol}^2=m_2^2-m_1^2$ and $\Delta m_{atm}^2=m_2^2-m_3^2$ we get an explicit relationship between  $\Delta m_{sol}^2$ and $\Delta m_{atm}^2$ as \bea
\Delta m_{atm}^2=\frac{1}{2}\frac{\Delta m_{sol}^2}{\sqrt{1+8p^2}}(4p^2-1)+\frac{\Delta m_{sol}^2}{2}
\eea
from which we obtain an  approximate range for $p$ through the experimental inputs of $3\sigma$ ranges.
In order to generate nonzero $\theta_{13}$ we invoke contribution from the charged lepton sector in the following way. We consider Altarelli-Ferugilo-Masina parametrization\cite{Altarelli:2004jb} for $U_{PMNS}$ which is written as $U_{PMNS}=U_l^\dagger U_{\nu}=\tilde{U}_l^\dagger diag(-e^{i\phi_1},e^{i\phi_2},1)U_\nu\times diag(1,e^{i\alpha},e^{i(\beta+\delta_{CP})})$,
 where $U_l$ diagonalizes the charged lepton mass matrix and $\tilde{U_l}$ follows usual CKM type parametrization as
 \bea
 \tilde{U_l}=\tilde{R}(\theta_{23})\tilde{R}(\theta_{13},\delta)\tilde{R}(\theta_{12})\label{ul}\nonumber\\
 \eea
 with
 \bea
 \tilde{R}(\theta_{23})=\begin{pmatrix}
 1&0&0\\0&\sqrt{1-\lambda_{23}^2}&\lambda_{23}\\0&-\lambda_{23}&\sqrt{1-\lambda_{23}^2}
 \end{pmatrix}, \nonumber \\ \tilde{R}(\theta_{13},\delta)=
 \begin{pmatrix}
 \sqrt{1-\lambda_{13}^2}&0&\lambda_{13}e^{i\delta}\\0&1&0\\-\lambda_{13}e^{-i\delta}&0&\sqrt{1-\lambda_{13}^2}
 \end{pmatrix}\nonumber \\  {\rm and}\hspace{.3cm}  \tilde{R}(\theta_{12})=
 \begin{pmatrix}
 \sqrt{1-\lambda_{12}^2}&\lambda_{12}&0\\-\lambda_{12}&\sqrt{1-\lambda_{12}^2}&0\\0&0&1
 \end{pmatrix}
 \eea
 along with $\lambda_{ij}=\sin \theta_{ij}$.\\
 As we are considering CKM type mixing matrix therefore, we expect small mixing in the charged lepton sector. Moreover the small value of reactor mixing angle also enforces the value of $\lambda$ to be small. The textures of the charged lepton mass matrices are presented in Table \ref{tu}
 \begin{table*}[ht]\centering
\caption{Texturs of the charged lepton mass matrix ($m_l$)} \label{tu}
\begin{center}
\begin{tabular}{|p{4.5cm}|p{4.5cm}|p{4.5cm}|}
\hline
\multicolumn{3}{|c|}{{\bf 4 zero textures of $m_l$ }}\\
\hline
$m_l^{12}=\begin{pmatrix}
\times&\times&0\\
\times&\times&0\\
0&0&\times\\
\end{pmatrix}$ & 
$m_l^{13}=\begin{pmatrix}
\times&0&\times\\
0&\times&0\\
\times&0&\times\\
\end{pmatrix}$ &
$m_l^{23}=\begin{pmatrix}
\times &0&0\\
0&\times&\times\\
0&\times&\times\\
\end{pmatrix}$\\
 \hline
\end{tabular}
\end{center}
\end{table*} \\
 where `$\times$' corresponds to some nonzero entries in $m_l$. Considering $|e_{\alpha=(e,\mu,\tau)}>$ to be the flavour eigenstate and $|e_i>$ the mass eigenstate of the charged leptons we address three  possible cases corresponding to the three textures of $m_l$ for  modifications of $U_{\nu}$. \\
 
 \noindent
{\bf{ Case I: $|e_{\tau}>^{flavour}=|e_i>^{mass}$, $m_l\Rightarrow m_l^{12}$}}\\

\noindent
In this case $U_{\nu}$ is modified by the 1-2 sector ($\tilde{R}(\theta_{12})$) of $U_l$ and the elements of $U_{PMNS}$ can be written as
\bea
U_{11}=-e^{i\phi_1}\sqrt{1-\lambda_{12}^2}c_{12}-\frac{1}{\sqrt{2}}\lambda_{12}e^{i\phi_2}s_{12}\nonumber\\
U_{12}=-e^{i\phi_1}\sqrt{1-\lambda_{12}^2}s_{12}+\frac{1}{\sqrt{2}}\lambda_{12}e^{i\phi_2}c_{12},\nonumber \\ U_{13}=-\frac{1}{\sqrt{2}}\lambda_{12}e^{i\phi_2}\nonumber\\
U_{22}=-\lambda_{12}e^{i\phi_1}s_{12}+\frac{1}{\sqrt{2}}\sqrt{1-\lambda_{12}^2}e^{i\phi_2}c_{12}\nonumber\\
U_{23}=-\frac{1}{\sqrt{2}}\sqrt{1-\lambda_{12}^2}e^{i\phi_2},U_{33}=\frac{1}{\sqrt{2}}
\eea
and hence the three mixing angles come out as
\bea
\sin \theta_{13}=|U_{13}|=\frac{\lambda_{12}}{\sqrt{2}}\nonumber\\
\tan \theta_{12}=\frac{|U_{12}|}{|U_{11}|}=\frac{s_{12}(s_{12}-\sqrt{2}\cos[\phi_1-\phi_2]c_{12}\lambda_{12})}{c_{12}(c_{12}+\sqrt{2}\cos[\phi_1-\phi_2]s_{12}\lambda_{12})}\nonumber\\
\tan \theta_{23}=\frac{|U_{23}|}{|U_{33}|}=\sqrt{1-\lambda_{12}^2}.
\eea
The measure of CP violation $J_{CP}$ can be written in terms of the mixing matrix elements as 
\bea
J_{CP}=\frac{\sin(\phi_2-\phi_1)c_{12}s_{12}\lambda_{12}}{2\sqrt{2}}
\eea
and hence the Dirac CP phase $\delta_{CP}$ is obtained as
\bea
\sin \delta_{CP}=\frac{J_{CP}}{\Omega}
\eea 
with the definition of $\Omega$ as \bea \Omega= c_{12}^{\prime}c_{13}^{\prime 2}c_{23}^{\prime}s_{12}^{\prime}s_{13}^{\prime}s_{23}^{\prime} \label{omg}\eea 

where $s^{\prime}_{ij}\Rightarrow \sin \theta_{ij} $ and  $c^{\prime}_{ij}\Rightarrow \cos \theta_{ij}$ are the usual mixing parameters in the CKM part of $U_{PMNS}$ which is defined as 
\bea
U_{PMNS}=P_\alpha U_{CKM}P_M
\eea with $P_{\alpha}=diag(e^{i\alpha_1},e^{i\alpha_2},e^{i\alpha_3})$ as the unphysical phase matrix, $U_{CKM}=\tilde{U}_l^\dagger diag(-e^{i\phi_1},e^{i\phi_2},1)U_\nu$ and $P_{M}=diag(1,e^{i\alpha},e^{i(\beta+\delta_{CP})})$ as the Majorana phsae matrix.
We use other two rephasing invariant quantities to calculate the Majorana phases as\cite{Iizuka:2014mba} 
\bea
\alpha&=&arg(U_{11}^*U_{12})\nonumber \\
\beta&=&arg(U_{13}U_{11}^*)
\eea
and thereby the phases come out as 
\bea
\tan \alpha=\frac{\sqrt{2}\sin (\phi_2-\phi_1 )\lambda_{12}}{\sqrt{2}\cos(\phi_2-\phi_1)(c_{12}^2-s_{12}^2)\lambda_{12}-2c_{12}s_{12}}
\eea
and \bea
\tan \beta=\frac{\sin (\phi_2-\phi_1)c_{12}}{\sqrt{2}\cos(\phi_2-\phi_1)c_{12}+s_{12}\lambda_{12}}.
\eea
{\bf Case II: $|e_{\mu}>^{flavour}=|e_i>^{mass}$,  $m_l\Rightarrow m_l^{13}$} 

\noindent
In this case modification to $U_\nu$ originates from 1-3 sector ($\tilde{R}(\theta_{13},\delta)$) of $U_l$ and the elements of $U_{PMNS}$ can be written as
\bea
U_{11}=-e^{i\phi_1}\sqrt{1-\lambda_{13}^2}c_{12}-\frac{1}{\sqrt{2}}\lambda_{13}e^{i\delta}s_{12}\nonumber\\
U_{12}=-e^{i\phi_1}\sqrt{1-\lambda_{13}^2}s_{12}+\frac{1}{\sqrt{2}}\lambda_{13}e^{i\delta}c_{12},\nonumber \\U_{13}=-\frac{1}{\sqrt{2}}\lambda_{13}e^{i\delta}\nonumber\\
U_{22}=\frac{1}{\sqrt{2}}c_{12},
U_{23}=-\frac{1}{\sqrt{2}},U_{33}=\frac{1}{\sqrt{2}}\sqrt{1-\lambda_{13}^2}
\eea
and hence the three mixing angles come out as
\bea
\sin \theta_{13}=|U_{13}|=\frac{\lambda_{13}}{\sqrt{2}}\nonumber\\
\tan \theta_{12}=\frac{|U_{12}|}{|U_{11}|}=\frac{s_{12}(s_{12}-\sqrt{2}\cos[\delta-\phi_1]c_{12}\lambda_{13})}{c_{12}(c_{12}+\sqrt{2}\cos[\delta-\phi_1]s_{12}\lambda_{13})}\nonumber\\
\tan \theta_{23}=\frac{|U_{23}|}{|U_{33}|}=\frac{1}{\sqrt{1-\lambda_{13}^2}}.
\eea
Proceeding in the same way as discussed in {\bf Case I}, $J_{CP}$ can be written in terms of the mass matrix elements as 
\bea
J_{CP}=\frac{\sin(\phi_1-\delta)c_{12}s_{12}\lambda_{13}}{2\sqrt{2}}
\eea
and \bea
\sin \delta_{CP}=\frac{J_{CP}}{\Omega}
\eea
where $\Omega$ is already defined in Eqn (\ref{omg}). Finally the Majorana phases are calculated as
\bea
\tan \alpha=\frac{\sqrt{2}\sin (\delta-\phi_1 )\lambda_{13}}{\sqrt{2}\cos(\delta-\phi_1)(c_{12}^2-s_{12}^2)\lambda_{13}-2c_{12}s_{12}}
\eea
and \bea
\tan \beta=\frac{\sin (\delta-\phi_1)c_{12}}{\sqrt{2}\cos(\delta-\phi_1)c_{12}+s_{12}\lambda_{13}}.
\eea
{\bf Case III: $|e_{e}>^{flavour}=|e_i>^{mass}$,  $m_l\Rightarrow m_l^{23}$ } 

\noindent
For this texture of $m_l$ (alternatively $\tilde{R}(\theta_{23})$ as the mixing matrix) it is not possible to generate $\theta_{13}$, hence is not taken into account. 

We also consider the other two matrices $N^2$ and $N^3$ and obtained all the mixing angles and eigenvalues. However, from numerical estimation it is found that both the cases do not admit the present experimental data and hence discarded.\\

\subsubsection{All nondegenerate eigenvalues of $M_{RS}$}\label{sec2.2.2}
Taking three different 6 zero textures of $m_D(m_D^{1,2,3})$ and one 3 zero texture of $\mu (\mu^2)$ with the  $M_{RS}$ as 
\bea
M_{RS}=diag(M_1,M_2,M_3)
\eea
  we construct three different textures of $m_\nu$ using inverse seesaw formula and they lead to
\bea
m_\nu^1 = \begin{pmatrix}
0&yp&ypq\\yp&0&yq\\ypq&yq&0
\end{pmatrix},
m_\nu^2 = \begin{pmatrix}
0&yq&yp\\yq&0&ypq\\yp&ypq&0
\end{pmatrix},\nonumber \\
m_\nu^3 = \begin{pmatrix}
0&ypq&yq\\ypq&0&yq\\yq&yp&0
\end{pmatrix}\label{alld0}
\eea
where $p=M_2/M_1$ and $q=M_2/M_3$ and $y=\mu y_i^2/M_2^2$ for each $m_\nu^i$. Now in the basis where the charged lepton mass matrix is diagonal one can easily construct the effective $m_\nu$s as 
\bea
m_{\nu f}= U_l^\dagger m_\nu^i U_l^*
\eea
where $U_l$ is already defined in Section\ref{sec2.2.1}. Since we are considering three specific textures of the charged lepton mass matrices (Table \ref{tu}), therefore, for a given $m_\nu$ we can construct three $m_{\nu f}$ taking contribution from each sectors of the charged leptons. Hence, we have altogether 9 effective $m_{\nu f}$. We consistently denote them as $m_{\nu f ij}$ after getting correction from the `$ij$'th sector of $U_l$. We do not present explicit structures of all the mass matrices. However, numerical estimation for each viable matrix is presented in the next section. 
\section{Numerical analysis and phenomenological discussion}\label{s3}
\textbf{i) Two degenerate eigenvalues of $M_{RS}$  }

Before going into the details of the numerical analysis an important point is to be noted that except $\tan \theta_{23}$ the expressions for the physical parameters obtained in Case II are the same as that of the Case I if we replace $\lambda_{13}$ by $\lambda_{12}$ and $\delta$ by $\phi_2$ and therefore the numerical estimation for one case can be  automatically translated to the other. Therefore from now on in a generic way we rename  $\lambda_{12}$ and $\lambda_{13}$ as $\lambda$.\\
\begin{table}[!h]
 \caption{Input experimental values\cite{Forero:2014bxa}}\label{t3}
 \begin{center}
 \begin{tabular}{|c|c|}
\hline 
Quantity & 3$\sigma$ ranges \\ 
\hline 
$|\Delta m_{31}^2|$ (N)& 2.30$< \Delta m_{31}^2(10^3 eV^{-2})<2.64$ \\ 
\hline 
$|\Delta m_{31}^2|$ (I)& 2.20$< \Delta m_{31}^2(10^3 eV^{-2})<2.54$ \\ 
\hline 
 $\Delta m_{21}^2$& 7.11$< \Delta m_{21}^2(10^5 eV^{-2})<8.18$ \\ 
\hline 
$\theta_{12}$ & $31.8^o<\theta_{12}<37.8^o$ \\ 
\hline 
$\theta_{23}$ &  $39.4^o < \theta_{23}<53.1^o$  \\  
\hline 
$\theta_{13}$ &  $8^o < \theta_{13}< 9.4^o$  \\ 
\hline 
\end{tabular} 
\end{center} 
\end{table}
\noindent 
 We consider small mixing arises from  the charged lepton sector and accordingly written down the expressions for the physical parameters with the terms dominant in $\lambda$. Moreover, the smallness of $\theta_{13}$ automatically implies that the order of $\lambda$ should be of the  order of  Sine of the reactor mixing angle. Taking into account the neutrino oscillation global fit data presented in Table \ref{t3} we randomly vary $\lambda$ and $\phi_2-\phi_1$ within the ranges as $0<\lambda<0.3$ and $-180^o<\phi_2-\phi_1<180^o$ and scan the parameter space. It is seen that the matrices of type $N^2$ and $N^3$ are not phenomenologically viable (even after considering charged lepton contribution ) as far as the present neutrino oscillation data is concerned. For $N^1$ type matrix we plot in figure \ref{fig1} the variation of p Vs y, $\lambda$ Vs $\Phi_2-\Phi_1$ and $\lambda$ Vs $\theta_{13}$ and it is depicted from the plots that the parameters  y and p vary within the ranges as $0.00071<y<0.00087$ and $38<p<51$ which is presented in the extreme left of figure \ref{fig1}. The ranges of $\lambda$ and $\phi_2-\phi_1$ are obtained as $0.197<\lambda<0.231$ and $35.5^0<\phi_2-\phi_1<74^o$, $-35.5^0<\phi_2-\phi_1<-74^o$ as one can read from the middle plot of  figure \ref{fig1}. Now since $|U_{e3}|$ is directly proportional to $\lambda$ it is needless to say that there is a linear variation of $|U_{e3}|$ with $\lambda$ and is depicted in the last plot of figure \ref{fig1}. As $\tan \theta_{12}$ has a strong dependence on $\phi_2-\phi_1$ we also present the variation of $\theta_{12}$ with $\phi_2-\phi_1$ in the extreme right panel of figure \ref{fig2}. The atmospheric mixing angle $\theta_{23}$ doesn't deviate much from $45^o$. For Case I, $\theta_{23}$ is smaller than the bi-maximal value while for the Case II it is slightly enhanced and we plot them in the first two figures of figure \ref{fig2}. This is a distinguishable characteristic between the two cases. Now as the CP violation in $U_{PMNS}$ is solely controlled by the phases arising from the charged lepton sector therefore we expect a great dependency of $\delta_{CP}$ on $\phi_2-\phi_1$ ($\delta-\phi_1$ for Case II) and a correlation between the CP phases. We plot $\delta_{CP}$ with $\phi_2-\phi_1$ in the extreme left panel of figure \ref{fig3} while the correlation of the Majorana phases with $\delta_{CP}$ is shown in the other two figures of figure \ref{fig3}. The ranges of the Dirac CP phase $\delta_{CP}$ is obtained as $38^o<|\delta_{CP}|<85^o$ while the Majorana phases are constrained as $30^o<| \beta|<65^0$ and $8^o<|\alpha|<17^o$. The $J_{CP}$ value is obtained within the range as $0.017<|J_{CP}|<0.04$  as one can read from the extreme left plot of figure \ref{fig4}. The model predicts inverted hierarchy of the neutrino masses which is explicit from the second figure of figure \ref{fig4}. We also obtain a range on the sum of three light neutrino masses as 0.0953 eV $<\Sigma_i m_i<$ 0.1026 eV  and a range of $|m_{11}|$ as  0.03 eV $<|m_{11}|<$ 0.048 eV which are well bellow the present experimental upper bound 0.23 eV and 0.35 eV respectively\cite{Ade:2013zuv}.

Before closing the discussion we would like to mention that although charged lepton correction to $\mu\tau$ symmetric matrix is studied before \cite{Iizuka:2014mba,Kitabayashi:2013afa,Duarah:2012bd,Dev:2013esa}, here we consider a two parameter structure of a $\mu\tau$ symmetric matrix which is much more predictive than the previous ones. As for example in our model CP violation arises completely from the charged lepton sector as our mass matrix consist of two real parameters. Thus mixing in the charged lepton sector dictates a common origin of $\theta_{13}$ and the CP-violating phases. In our analysis the Dirac and Majorana phases are significantly correlated. Thus only the measurement of CP violating phases can challenge the viability of the present model \cite{Shimizu:2015tta}. With the recent hint of T2K, nearly maximal CP violation\cite{Samanta:2016nat} is also allowed here which in turn fixes the Majorana phases and thus the double beta decay parameter $|m_{11}|$. The allowed occurrence of inverted hierarchy puts a lower limit to $|m_{11}|$ as shown in figure \ref{fig5}. One can see a very narrow range of $|m_{11}|$ is allowed. Thus significant development  of the experiments like  GERDA and EXO  can test the viability of the model. Finally the constraint range of the sum of the light neutrino masses is also a major result of the analysis as $\Sigma_im_i\sim$ 0.1 eV at $4\sigma$ level is expected to be probed by the future astrophysical experiments.

\noindent
\textbf{ii) All nondegenerate eigenvalues of $M_{RS}$}

In this category there are 9 structures of effective $m_\nu$ matrices. We diagonalize them through a direct  diagonalization procedure \cite{Adhikary:2013bma} and calculate the eigenvalues, mixing angles. It is seen that the matrices $m^1_{\nu f23}$, $m^2_{\nu f23}$ and $m^3_{\nu f23}$ are phenomenologically ruled out. To be more specific one needs $\lambda_{23}\gg1$ which is not be the case. Proceeding in the same way as that of the previous section we estimate the ranges of $J_{CP}$, $\delta_{CP}$, $\alpha,\beta$, $|m_{11}|$ and $|\Sigma_i m_i|$  for the survived matrices. The hierarchy of the neutrino masses for all the cases is inverted. The predictions of the viable  matrices are listed in Table \ref{t6}. In figure \ref{fig6} we plot the lightest eigenvalue with $|m_{11}|$. \\
Before concluding this section we would like to mention that the charged lepton correction to the matrices given in Eqn. (\ref{alld0}) (with all diagonal entries zero) are also studied in Ref.\cite{Ludl:2014axa}. Particularly the classes $4_4$ and $3_1$\cite{Ludl:2014axa}  respectively resemble $m_l$ and $m_\nu$ matrices  considered here in the present work. However, in  Ref.\cite{Ludl:2014axa}  these cases are categorized as less predictive due to large number of parameters (10 real parameters) and hence the results are not presented. However in the present work, those cases contain less number of parameters (7 real parameters) since the structure of $M_{RS}$ matrix is flavour diagonal ($p$ and $q$ parameters defined in Section{\ref{sec2.2.2}} are real) and we estimate the prediction for these cases regarding $|m_{11}|$, $\Sigma_im_i$, $\delta_{CP}$ etc.

\noindent
Some concluding remarks  regarding predictions of the present scheme:\\
\noindent  
  1. We see the hierarchy of neutrino mass predicted in all the cases  is inverted. This can be testified in the near future through the combined analysis of NO$\nu$A and T2K\cite{Ieki:2014bca} experimental data with the aid of the knowledge of precise $\theta_{13}$. The Majorana phases do not appear in $neutrino\rightarrow neutrino$ oscillation  experiments, however, they may appear in $neutrino\rightarrow$ $anti$ $neutrino$ oscillation experiments. Although these experiments are difficult to design, however, in an optimistic point of view we expect the prediction for the Majorana phases in this model will also be tested in  future experiments.\\
\noindent  
2. The predicted value of $\Sigma_i m_i$ obtained in the present model could also be tested in the near future through more precise estimation of $\Sigma_im_i$ due to a combined analysis using PLANCK data\cite{Ade:2013zuv} and other cosmological and astrophysical experiments\cite{Lesgourgues:2014zoa,1475-7516-2013-01-026}. The value of $\Sigma_im_i\sim$ 0.1 eV at the 4$\sigma$ level could be probed through such analysis for the inverted ordering of the neutrino masses. \\
3. The Dirac CP phase $\delta_{CP}$ predicted in this work will be tested at NO$\nu$A, T2K and DUNE\cite{Samanta:2016wca} in near future.
\section{Summary and Conclusion}\label{s4}
 Within the framework of inverse seesaw, we study the phenomenology of maximal zero textures and cyclic symmetry in the neutrino  matrices.  We  adopt two different schemes to accommodate the present oscillation data. In the first approach we consider explicit cyclic  symmetry on the relevant  fields which leads to degenerate eigenvalues. To remove the degeneracy in the eigenvalues we incorporate explicit symmetry breaking term in the Lagrangian. It is seen that even after considering the charged lepton correction from the cyclic symmetry invariant $m_l$, present oscillation data can not be explained and hence the first approach is discarded. In the second one, we concentrate on the same form of the neutrino mass matrices which can be realized through an effective residual symmetry of some bigger symmetry group in the Lagrangian, in which cyclic symmetry in the charged lepton sector is broken after spontaneous breaking of the bigger group. Further we exemplify a toy model with softly broken $A_4$ symmetry to realize one of the combinatorial structure of effective $m_\nu$. To fit the oscillation data charged lepton correction from different sectors of $U_l$ is considered along with a soft breaking term in $M_{RS}$ which removes the degeneracy in masses. Each cases predict a highly constrained ranges of CP violating phases, $|m_{11}|$ and  $\Sigma_i m_i$ along with inverted ordering of the neutrino masses. \\
 \textbf{Acknowledgement}\\
Authors acknowledge Department of Atomic Energy (DAE), Government of India, for financial support.\\

 \begin{table*}[ht]\centering
\caption{Predictions of the viable matrices.}\label{t6}
\begin{center}
\begin{tabular}{|p{1.2cm}|p{1.9cm}|p{1.9cm}|p{1.9cm}|p{1.9cm}|p{1.9cm}|p{2.3cm}|}
\cline{2-7}
\multicolumn{1}{c|}{}  & \multicolumn{6}{c|}{ Six predicted quantities } \\
\hline
&$|\delta_{CP}|$ (deg.) &$|\alpha|$ (deg.)&$|\beta|$ (deg.) & $|J_{CP}|$&$\Sigma_im_i$ (eV) &$|m_{11}|$ (eV)\\
\hline
$m^1_{\nu f12}$&$100-23$ &$-80-12$&$-63-23$ &$0.01-0.04$&$0.09-0.12$&$0.026-0.048$\\
\hline
$m^2_{\nu f12}$&$98-34$ &$92-18$&$78-0$ & $0.015-0.38$&$0.07-0.108$&$0.029-0.049$\\
\hline
$m^3_{\nu f12}$&$88-0$ &$71-37$&$62-35$ & $0.012-0.036$&$0.07-0.1$&$0.029-0.048$\\
\hline
$m^1_{\nu f13}$&$100-20$ &$85-10$&$60-20$ &$0.01-0.04$&$0.09-0.14$&$0.031-0.05$\\
\hline
$m^2_{\nu f13}$&$94-35$ &$100-18$&$81-14$ & $0.012-0.038$&$0.07-0.132$&$0.032-0.049$\\
\hline
$m^3_{\nu f13}$&$102-17$ &$82-26$&$62-21$ & $0.017-0.04$&$0.08-0.15$&$0.028-0.048$\\
\hline
\end{tabular}
\end{center}
\end{table*}
\begin{figure*}[ht]\centering
\includegraphics[scale=.3]{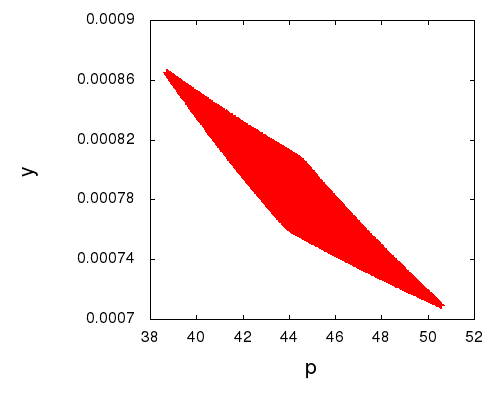} \includegraphics[scale=.3]{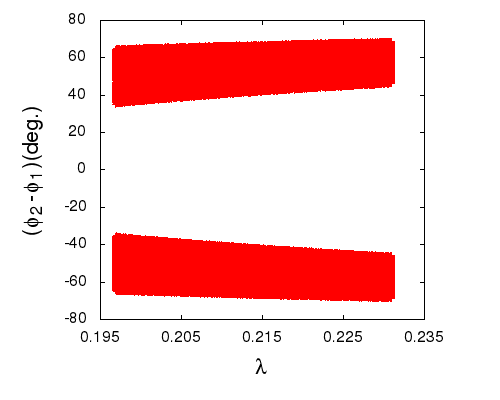}\includegraphics[scale=.3]{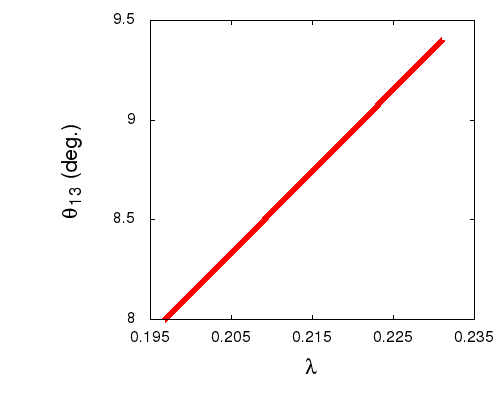}
\caption{(colour online) Correlation plots: Extreme left plot represents y Vs p while the middle one shows  $\lambda$ Vs $\phi_2-\phi_1$ for Case I. For Case II we get the same plot just by replacing $\phi_2-\phi_1$ with $\delta-\phi_1$ and finally the plot in the extreme right shows the variation of $\lambda$ with $\theta_{13}$ for both the cases.}\label{fig1}
\end{figure*}

\begin{figure*}[!htbp]\centering
 \includegraphics[scale=.3]{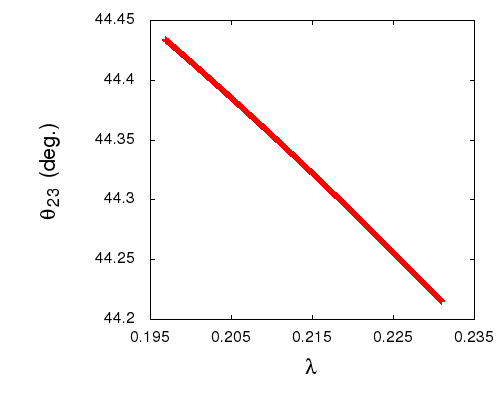}  \includegraphics[scale=.3]{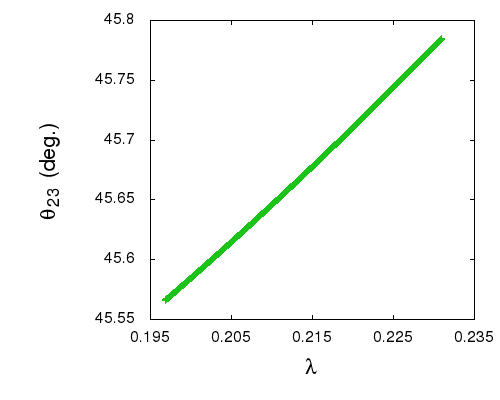}\includegraphics[scale=.3]{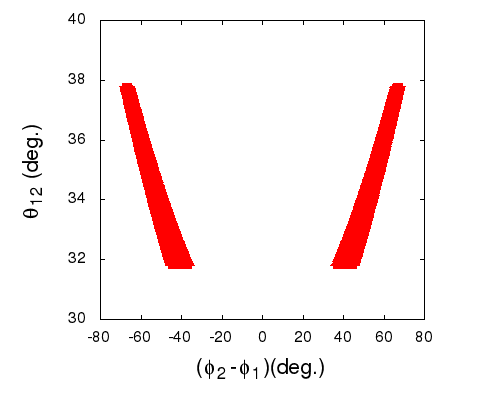}
 \caption{(colour online) The first figure (the red line) shows the variation of the atmospheric mixing angle ($\theta_{23}$) with $\lambda$ for Case I while the second one (the green line) shows the same for Case II. The last one represents the correlation between $\theta_{12}$ and $\phi_2-\phi_1$ for case I. We get the same plot for case II by replacing $\phi_2-\phi_1$ with $\delta-\phi_1$. }\label{fig2}
\end{figure*}

\begin{figure*}[!htbp]\centering
 \includegraphics[scale=.3]{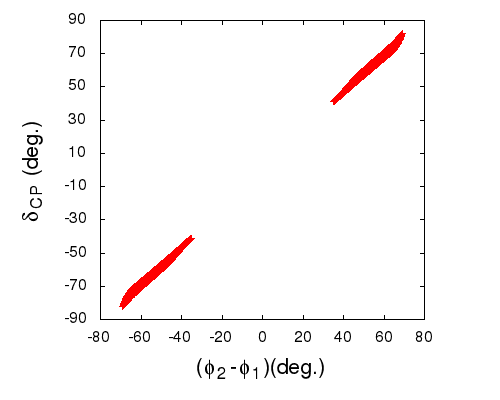} \includegraphics[scale=.3]{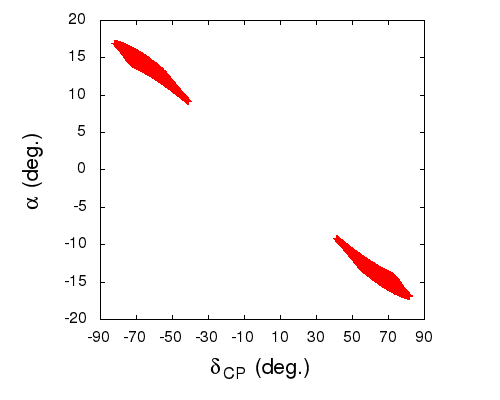} \includegraphics[scale=.3]{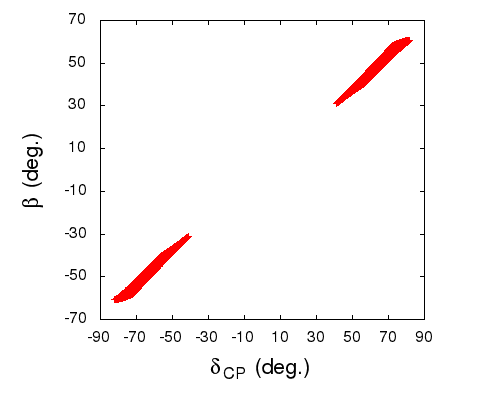}
 \caption{(colour online) The plot in the extreme left side shows the variation of $\phi_2-\phi_1$ with $\delta_{CP}$ for Case I and we get the same plot for Case II by replacing $\phi_2-\phi_1$ with $\delta-\phi_1$ while the other two plots show the correlation between the Majorana phases with $\delta_{CP}$ and are same for both the cases.}\label{fig3}
\end{figure*}
\begin{figure*}[!htbp]\centering
 \includegraphics[scale=.3]{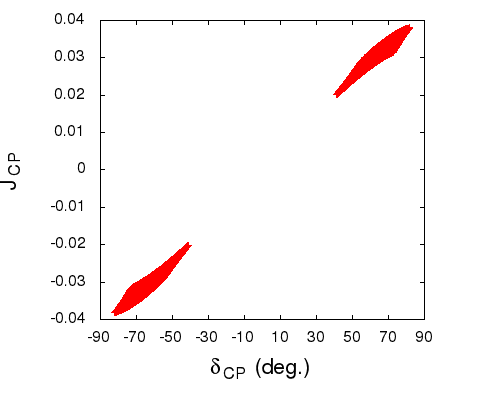} \includegraphics[scale=.3]{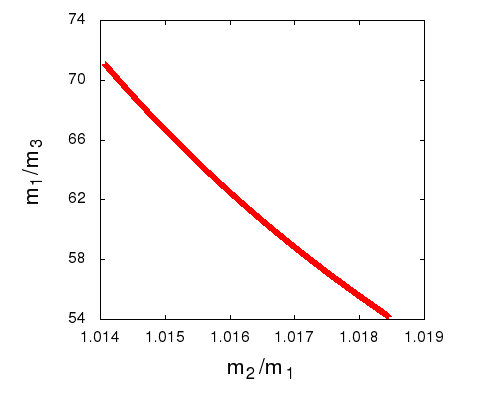}\includegraphics[scale=.3]{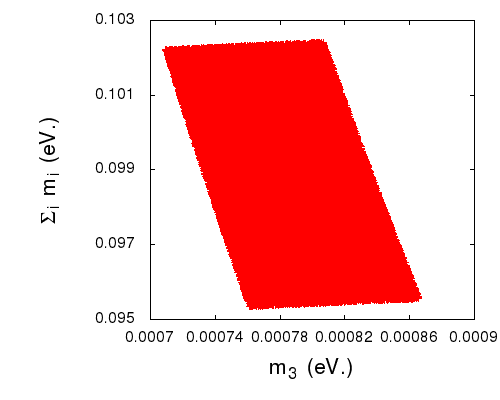}
 \caption{(colour online) The first one shows the variation of $J_{CP}$ with $\delta_{CP}$, the second one stands for the inverted hierarchy of neutrino masses and last one shows a correlation between $\Sigma_i m_i$ with $m_3$ and all the plots presented in this figure are same for both the cases.}\label{fig4}
\end{figure*}
\begin{figure*}[!htbp]\centering
\begin{center}
 \includegraphics[scale=.3]{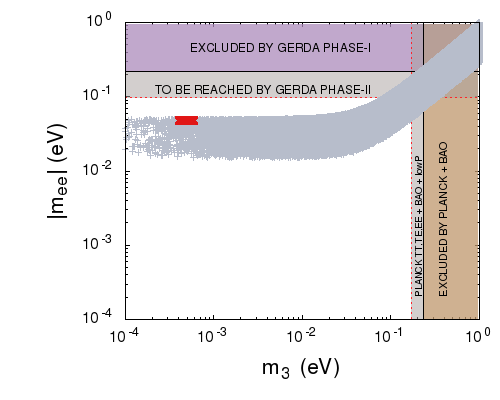} 
 \caption{(colour online) Lightest eigenvalue ($m_3$) Vs $|m_{11}|$ plot. The gray band shows the range of $|m_{11}|$ allowed by the present oscillation data with all the CP phases within the range $0-2\pi$. The small red coloured band is allowed in our model.}\label{fig5}
 \end{center}
\end{figure*}
\begin{figure*}[ht]\centering
\includegraphics[scale=.3]{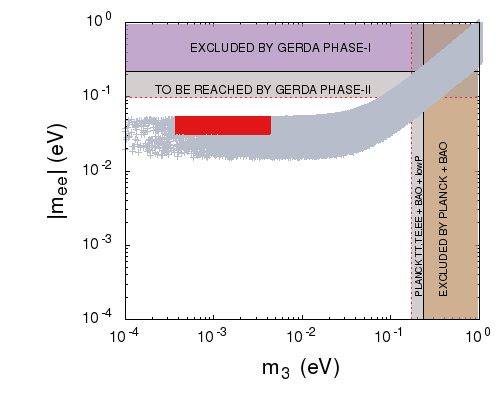} \includegraphics[scale=.3]{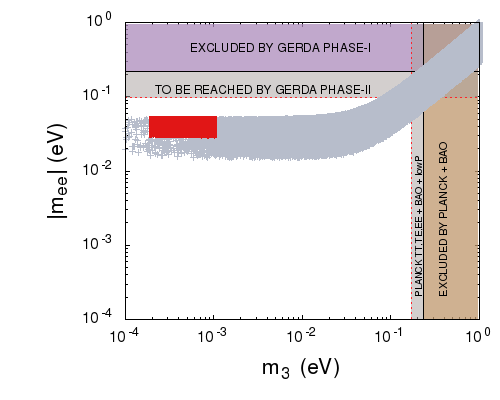}\includegraphics[scale=.3]{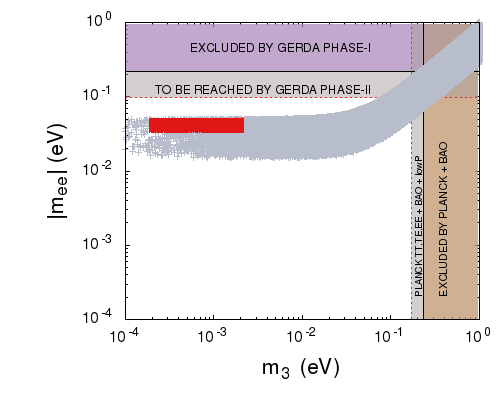}\\
\includegraphics[scale=.3]{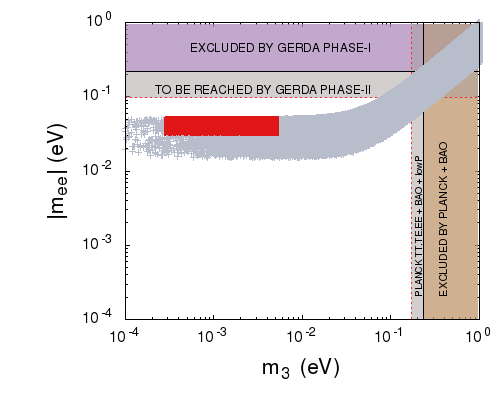} \includegraphics[scale=.3]{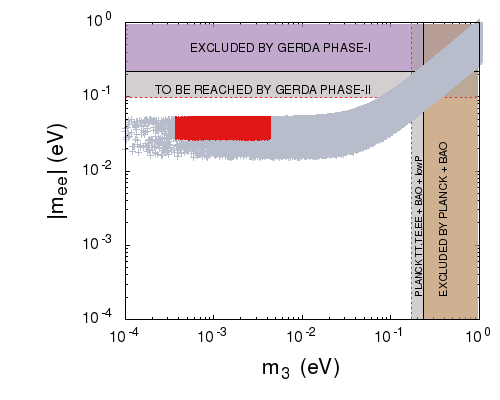}\includegraphics[scale=.3]{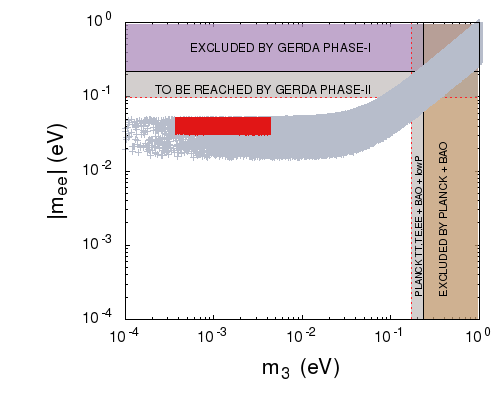}
\caption{(colour online) Lightest eigenvalue ($m_3$) Vs $|m_{11}|$ plot: For all nondegenerate eigenvalues of $M_{RS}$. The first three figures of the first row are shown for the matrices $m^{i(=1,2,3)}_{\nu f12}$ and the figures in the second row are shown for the matrices $m^{i(=1,2,3)}_{\nu f13}.$ }\label{fig6}
\end{figure*}

{}

\end{document}